\documentclass[sn-mathphys-num,a4paper]{sn-jnl}

\usepackage{mathtools}
\usepackage{array}
\usepackage{tabularx}
\usepackage{graphicx}
\usepackage{dcolumn}
\usepackage{textcomp}
\usepackage{bm}
\usepackage{amsmath}
\usepackage{color}
\usepackage{siunitx}
\usepackage{wasysym}
\usepackage{float}
\usepackage{array,url,kantlipsum}
\usepackage{verbatim}
\usepackage[dvipsnames]{xcolor}
\usepackage{CJK}
\usepackage{eurosym}
\usepackage{caption}
\usepackage{xcolor}
\usepackage{subcaption}
\newcolumntype{P}[1]{>{\centering\arraybackslash}p{#1}}
\usepackage{array}
\usepackage{inputenc}
\usepackage[section]{placeins}
\usepackage{amssymb}
\usepackage{siunitx}  
\usepackage{multirow}%
\usepackage{amsfonts}%
\usepackage{amsthm}%
\usepackage{mathrsfs}%
\usepackage[title]{appendix}%
\usepackage{manyfoot}%
\usepackage{booktabs}%
\usepackage{algorithm}%
\usepackage{algorithmicx}%
\usepackage{algpseudocode}%
\usepackage{listings}%
\usepackage{makecell}

\begin{document}

\date{\today}
\title[Article Title]{Human Cardiac Measurements with Diamond Magnetometers }

\author[1,2]{Muhib Omar}

\author[3]{Magnus Benke}

\author*[1,2]{Shaowen Zhang}\email{shzhang@uni-mainz.de}

\author*[3]{Jixing Zhang}\email{jixing.zhang@pi3.uni-stuttgart.de}

\author[3]{Michael Kuebler}

\author[1,2]{Pouya Sharbati}

\author[1,2]{Ara Rahimpour}

\author[1,2]{Arno Gück}

\author[14,15]{Maryna Kapitonova}

\author[11]{Devyani Kadam}

\author[5,6]{Carlos René Izquierdo Geiser}

\author[7]{Jens Haller}

\author[7]{Arno Trautmann}

\author[7]{Katharina Jag-Lauber}

\author[7]{Robert Rölver}

\author[8]{Thanh-Duc Nguyen}

\author[8]{Leonardo Gizzi}

\author[3]{ Michelle Schweizer}

\author[16]{ Mena Abdelsayed}

\author[10]{Ingo Wickenbrock}

\author[9]{Andrew M. Edmonds}

\author[9]{Matthew Markham}

\author[11]{Peter A. Koss}

\author[12]{Oliver Schnell}

\author[5,6]{Ulrich G. Hofmann}

\author[14,15]{Tonio Ball}

\author[5,13]{Jürgen Beck}
 
\author[1,2,4]{Dmitry Budker}

\author[3]{Joerg Wrachtrup}

\author[1,2]{Arne Wickenbrock}

\affil[1]{ \orgname{Helmholtz-Institut Mainz, GSI Helmholtzzentrum f{\"u}r Schwerionenforschung}, \orgaddress{\street{Staudingerweg 18}, \city{Mainz}, \postcode{55128}, \country{Germany}}}

\affil[2]{ \orgname{Johannes Gutenberg-Universit{\"a}t Mainz}, \orgaddress{\street{Saarstr. 21}, \city{Mainz}, \postcode{55122}, \country{Germany}}}

\affil[3]{\orgdiv{Zentrum für Angewandte Quantumtechnologie (ZAQuant)}, \orgname{University of Stuttgart}, \orgaddress{\street{Allmandring 13}, \city{Stuttgart}, \postcode{70569}, \country{Germany}}}

\affil[4]{\orgdiv{Department of Physics}, \orgname{University of California}, \orgaddress{\street{366 Physics North, MC 7300}, \city{Berkeley}, \postcode{94720}, \state{California}, \country{USA}}}

\affil[5]{\orgdiv{Medical Faculty}, \orgname{University of Freiburg}, \orgaddress{\street{Hugstetter Str. 55}, \city{Freiburg im Breisgau}, \postcode{79106}, \country{Germany}}}

\affil[6]{\orgdiv{Department of Neurosurgery, Medical center,}  \orgname{University of Freiburg}, \orgaddress{\street{Breisacher Str. 64}, \city{Freiburg im Breisgau}, \postcode{79106}, \country{Germany}}}

\affil[7]{\orgdiv{Q.ANT GmbH}, \orgaddress{\street{Handwerkstraße 29}, \city{Stuttgart}, \postcode{70565}, \country{Germany}}}

\affil[8]{\orgdiv{Fraunhofer Institute for Production Engineering and Automation}, \orgaddress{\street{Nobelstraße 12}, \city{Stuttgart}, \postcode{70569}, \country{Germany}}}

\affil[9]{\orgdiv{Element Six Global Innovation Centre}, \orgaddress{\street{Fermi Avenue, Harwell Oxford, Didcot}, \city{Oxfordshire}, \postcode{OX11 0QR}, \country{United Kingdom}}}

\affil[10]{\orgdiv{Medizinische Klinik I, Abteilung für Kardiologie, Elektrophysiologie, Pneumologie und konservative
Intensivmedizin, Klinikum Lünen}, \orgaddress{\street{Altstadtstraße 23}, \city{Lünen}, \postcode{44534}, \country{Germany}}}

\affil[11]{\orgdiv{Fraunhofer Institute for Physical Measurement Techniques (IPM)}, \orgaddress{\street{Georges-Köhler-Allee 301}, \city{Freiburg}, \postcode{79110}, \country{Germany}}}

\affil[12]{\orgdiv{Department of Neurosurgery, Uniklinikum Erlangen, Friedrich-Alexander-Universität Erlangen-Nürnberg}, \orgaddress{\street{Hugenottenpl. 6}, \city{ Erlangen}, \postcode{91054}, \country{Germany}}}

\affil[13]{\orgdiv{Center for Advanced Surgical Tissue Analysis (CAST), Faculty of Medicine, University of Freiburg}\city{, Freiburg}, \country{Germany}}

\affil[14]{\orgdiv{ Neuromedical AI Lab, Department of Neurosurgery, Medical Center—University of Freiburg, Faculty of Medicine, University of Freiburg}\city{, Freiburg}, \country{Germany}}

\affil[15]{\orgdiv{Institute for Machine-Brain Interfacing Technology, University of Freiburg}\city{, Freiburg}, \country{Germany}}

\affil[16]{\orgdiv{Lankenau Institute for Medical Research}\city{, Pennsylvania}, \country{USA}}

\abstract{We demonstrate direct, non-invasive and non-contact detection of human cardiac magnetic signals using quantum sensors based on nitrogen-vacancy (NV) centers in diamond. Three configurations were employed recording magnetocardiography (MCG) signals in various shielded and unshielded environments. The signals were averaged over a few hundreds up to several thousands of heart beats to detect the MCG traces. The compact room-temperature NV sensors exhibit sensitivities of 6–26\,pT/$\sqrt{\text{Hz}}$ with active sensing volumes below 0.5\,\,$\text{mm}^3$, defining the performance level of the demonstrated MCG measurements. While the present signals are obtained by averaging, this performance already indicates a clear path toward single-shot MCG sensing. To move beyond shielded environments toward practical clinical use, strong noise suppression is required. To this end, we implement NV-based gradiometry and achieve efficient common-mode noise rejection, enabled by the intrinsically small sensing volume of NV sensors. Together, these multi-platform results obtained across diverse magnetic environments provide a solid foundation for translating quantum sensors into human medical diagnostics such as MCG and magnetoencephalography (MEG).}

\maketitle

\section{Introduction}
\label{Introduction}
Weak magnetic fields generated by neuronal and muscular activity, especially cardiac magnetic fields, carry rich physiological information. Magnetocardiography (MCG) can be used to diagnose conditions such as arrhythmias and coronary artery disease\,\cite{reviewmcg,Fenici2005ClinicalMCG}. 

Electrocardiography (ECG) and MCG are both based on cardiac electrical activity, with the former recording electric potential differences and the latter recording magnetic fields, reflecting the same physiological processes of depolarization and repolarization. ECG signals originated from action potentials in the heart and can be recorded on the body surface through electrodes. This technique is mature, low-cost, and widely available, making it the clinical gold standard for routine diagnosis. However, ECG signals are strongly influenced by the conductivity of body tissues, resulting in limited spatial resolution while the necessity to contact the skin precludes application when such contact is not possible (e.g. burn victims). In contrast, MCG signals arise from the magnetic fields generated by cardiac currents and can be detected by highly sensitive magnetometers. As a non-contact technique, MCG signals are minimally affected by tissue conductivity, enabling the detection of subtle abnormalities that ECG may miss and supporting precise localization of pathological sources\,\cite{reviewmcg}. Historically, MCG relied on SQUIDs and later on optically-pumped magnetometers based on vapor cells (OPMs). Other magnetometers, such as fluxgate sensors and magnetoresistive sensors, have also been demonstrated to detect MCG signals\,\cite{Oogane2021SubpT,Karo2016First36Channel}. Although MCG offers high clinical potential, the cardiac magnetic signals are inherently weak and the measurements are highly susceptible to environmental magnetic noise, necessitating stringent magnetic shielding. In addition, the high cost of equipment and the environmental dependence of present-day sensors have limited its widespread clinical adoption. Additionally, the ECG and MCG techniques are complementary (see Table\,\ref{tab1}), and their integration may substantially improve early diagnosis and precise localization of cardiac diseases.
Current MCG technologies are predominantly based on superconducting quantum interference devices (SQUIDs), see for example \,\cite{mcgsquid},  which require cryogenic cooling and thus hinder widespread clinical deployment. In unshielded environments OPMs have been implemented as an alternative\,\cite{unshieldedopm}. But following this approach requires facing the challenges due to heading errors, dead zones, and high operating temperatures that complicate contact-based measurements\,\cite{Kiehl2024heading,Tian2024DeadZone}. Therefore, development of miniaturized, room-temperature MCG technologies capable of providing full vector information in unshielded environments remains desirable.

One such pathway is based on nitrogen-vacancy (NV) centers in diamond.  During the past two decades, NV sensors have been widely used to measure various physical quantities, including temperature \,\cite{temperature2}, strain \cite{strainimaging}, electric fields \cite{electric1}, magnetic fields \cite{barryensing}, and rotation \cite{PhysRevLett.126.197702}. Compared with OPMs, NV magnetometers feature a much smaller sensing volume, simplifying gradiometer based, vector-resolved MCG measurements under unshielded conditions. Their comparatively high volume sensitivity\,\cite{2019BarryNVReview} makes them credible competitors to OPMs when combined with flux concentrators\,\cite{fescenko2020diamond}. The comparably lower sensitivity may then be an advantage, since the intrinsic noise of the flux concentrator might not become a limiting factor\,\cite{Griffith2009Miniature}. They also offer fast initialization, excellent biocompatibility, and stable operation over a broad temperature range showing great potential for biomedical sensing. NV magnetometers have been demonstrated in animal studies for both invasive \,\cite{keigomcg} and non-invasive MCG detection using flux concentrators \,\cite{fluxmcg}. 

In this study, we detected human cardiac magnetic signals using NV magnetometers independently developed by three groups, Johannes Gutenberg-University Mainz (JGU), University of Stuttgart (ZAQuant) and the quantum technology start-up Q.ANT GmbH (Q.ANT). These results represent a critical step toward the real-world biomedical application of quantum magnetometers and demonstrate that NV diamond magnetometry has now reached a stage where meaningful biomagnetic signals can be measured under various experimental conditions. This progress allows to quantify the needed improvement in signal-to-noise-ratio (SNR) and positions NV sensing as being competitive with existing technologies while promising to offer improved accessibility and practicality for both clinical and home use.

\section{Results}
\label{Results}
This section describes the three integrated NV magnetometers and the corresponding MCG measurement procedures, covering both zero-bias and finite bias-field configurations operated in shielded, partially shielded, and unshielded environments. It further reports a cross-validation of the NV-based MCG signals using a commercial OPM sensor array, and concludes with gradiometric noise-suppression results that demonstrate the feasibility of NV magnetometry in realistic, noisy, and unshielded settings.

\subsection{MCG with NV-based magnetometers}
\label{MCG based on (zero) bias field NV magnetometer}

 \begin{figure}
    \centering
    \includegraphics[width=1\columnwidth]{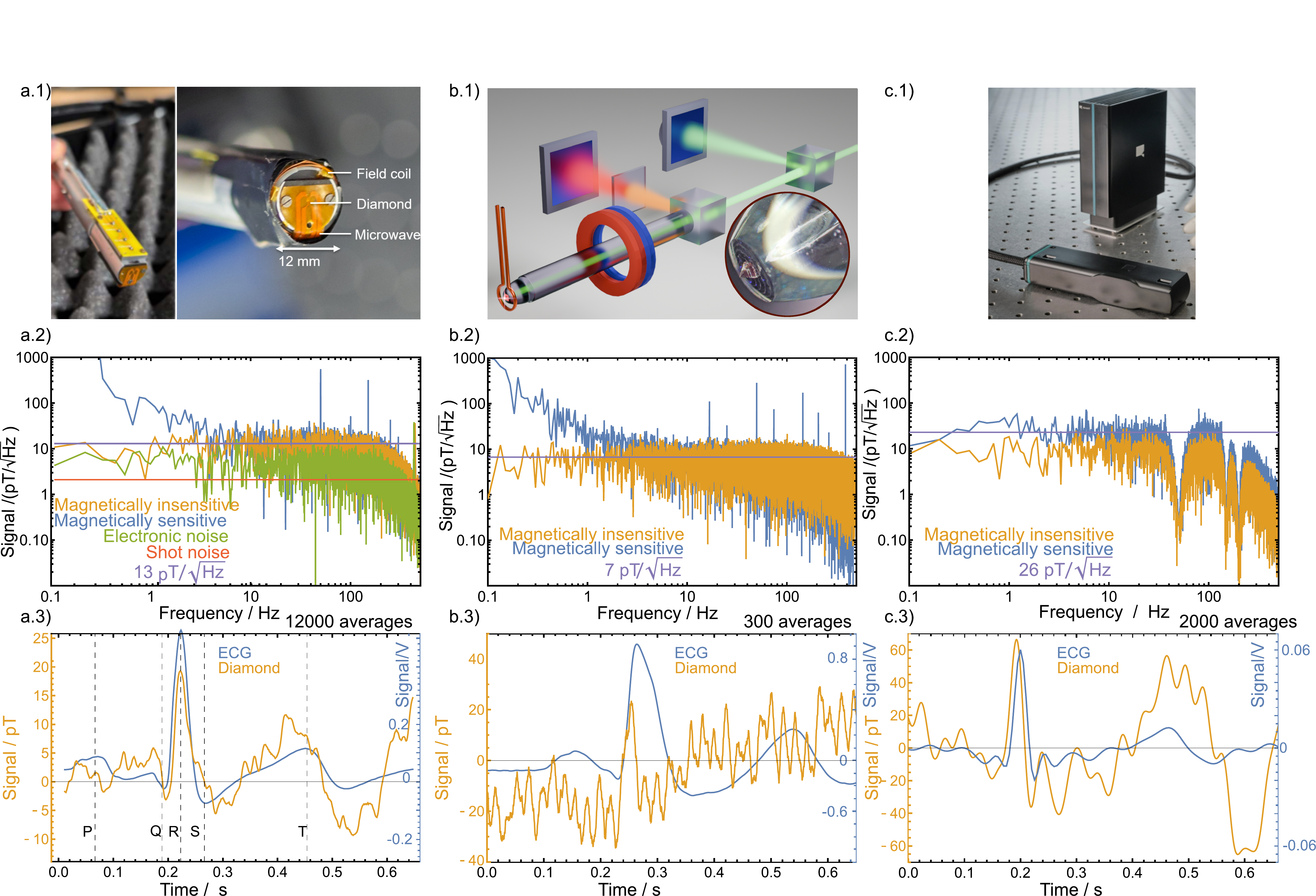}
    \caption{\textbf{Column (a)}  JGU: a.1):  Diamond sensor head dimensions.
    a.2): Amplitude spectral density with a baseline of 13\,pT/$\sqrt{\mathrm{Hz}}$, measured inside the shielded room of the Helmholtz Institute Mainz (JGU). a.3): ECG trace and diamond MCG signal (12000 averages) with indications of the main components of the ECG, P wave, QRS complex and T wave.
    \textbf{Column (b)} ZAQuant:
    b.1): Sensor schematic.
    b.2): Measured amplitude spectral densities inside the partially shielded environment, showing a noise baseline of 7\,pT/$\sqrt{\mathrm{Hz}}$. b.3): ECG trace and averaged diamond MCG signal (300 averages). 
    \textbf{Column (c)} Q.ANT:
   a.1): Sensor design. 
   a.2): Amplitude spectral density with a baseline of 26\,pT/$\sqrt{\mathrm{Hz}}$; filters were applied during acquisition. 
   \a.3): ECG trace and diamond MCG signal (2000 averages).}
    \label{fig:1}
\end{figure}

We first describe the JGU MCG measurements conducted at the Helmholtz Institute Mainz within its multilayer magnetically shielded room using the zero-bias-field NV magnetometer. Afterwards, the MCG measurements realized with bias-field NV magnetometers at the University of Stuttgart and at Q.ANT GmbH are introduced. Schematics of the sensors, their noise characteristics and the MCG measurements for all sensors can be seen in Fig.\,\ref{fig:1}. 

To record human cardiac magnetic signals in Mainz, we employed a fiberized NV diamond magnetometer designed for the use as a transportable endoscope. The sensor is operated without a magnetic bias field. The sensing protocol employing an oscillating magnetic field to recover magnetic sensitivity without a dc bias magnet is described in \cite{zerohuijie,zulfpaper}. Zero-field shielded environments are straightforward to create and minimize bias field fluctuations and magnetic field gradients\,\cite{Altarev2014MagneticallyShieldedRoom}. This makes them compatible with existing setups for optically pumped magnetometers (OPMs) used to validate the MCG detection (see section\,\ref{Validating the MCG signature}). The sensing diamond\,\cite{diamondoptics} was a truncated pyramid with 0.18\,mm height and a 0.5\,mm x 0.5\,mm base. A magnetic sensitive noise level of 13\,pT/$\sqrt{\mathrm{Hz}}$ (averaged between 20\,Hz and 30\,Hz) was achieved in the magnetically shielded room [comp. Fig.\,\ref{fig:1} a.2)]. This value was based on individual time-series measurements of 9\,s duration, processed using a Hann window prior to Fourier transformation. The amplitude spectrum was then divided by the square root of the effective noise bandwidth\,\cite{fft} to obtain the amplitude spectral density. Magnetically insensitive traces as well as electronic-noise traces were also recorded. The magnetically insensitive noise floor indicated residual laser noise as the limiting noise source. A detailed discussion of the noise-floor characteristics and sources can be found in \cite{zulfpaper}.  The endoscopic sensor design is described in detail in\,\cite{zulfpaper} and shown in Fig.\,\ref{fig:1}\,a) and in Fig.\,\ref{fig:2}\,c) with the sensor head components. 
 During data collection, the sensor was positioned approximately 1\,cm away from a seated subject's chest, leaving sufficient space for breathing. Data were recorded in 22\,min intervals. Electrodes were placed on the subject's back to record reference ECGs used as a trigger to average the data across multiple cardiac cycles to account for heart-rate variability \cite{HeartRateVariability}. Averaging was necessary since magnetic field recordings did not reveal the MCG signal in the NV trace in real-time. 

For scalable bedside MCG applications in unshielded settings, bias-field NV magnetometers could be advantageous. In addition to requiring a stable bias magnetic field, they must also suppress ambient magnetic noise, which poses a major challenge in real-world environments. The sensors from ZAQuant and Q.ANT followed this route. These sensors are described in Fig.\,\ref{fig:1} column b) and c). The sensitivity analysis was carried out analogously to the zero-field NV sensor, the only difference being that the time trace durations were 10\,s (Q.ANT) and 60\,s (ZAQuant). Sensitivities of 7\,$\mathrm{pT}/\sqrt{\mathrm{Hz}}$ (ZAQuant) and 26\,$\mathrm{pT}/\sqrt{\mathrm{Hz}}$ (Q.ANT) were determined. For the MCG measurements at ZAQuant the subject was sitting \SI{\sim 1}{cm} away from the sensor head. The Q.ANT measurements were performed with a subject lying directly below the sensor. The magnetic signal was recorded simultaneously with ECG to enable averaging. The MCG data from the ZAQuant measurements were filtered with narrow band-stop filters around \SI{50}{Hz}, \SI{60}{Hz} and their harmonics to reduce the noise of these common interferences. The Q.ANT MCG data was subjected to four notch filters around \qtylist{8.8;14.7;16.7;50}{Hz} with a quality factor of \num{30}, a Butterworth low-pass filter with \SI{30}{Hz} and a high-pass filter with \SI{8}{Hz} before averaging. 
Recordings showed MCG results for each of the sensors. The amount of averages for the displayed data were 12000, 300, and 2000 for JGU, ZAQuant and Q.ANT, respectively. 
	
\subsection{Validating the MCG signature}
\label{Validating the MCG signature}
The JGU MCG signal was visible with the diamond sensor after averaging, triggering with the ECG signal from an electrode on the subject's back. To be certain that the electrical signal from the heart was not accidentally picked up by the measurement electronics of the diamond sensor, diamond MCG was recorded at two positions with different MCG peak polarity with the same ECG setting. This ruled out electronic cross talk.

\begin{figure}[h]
    \centering
    \includegraphics[width=1\columnwidth]{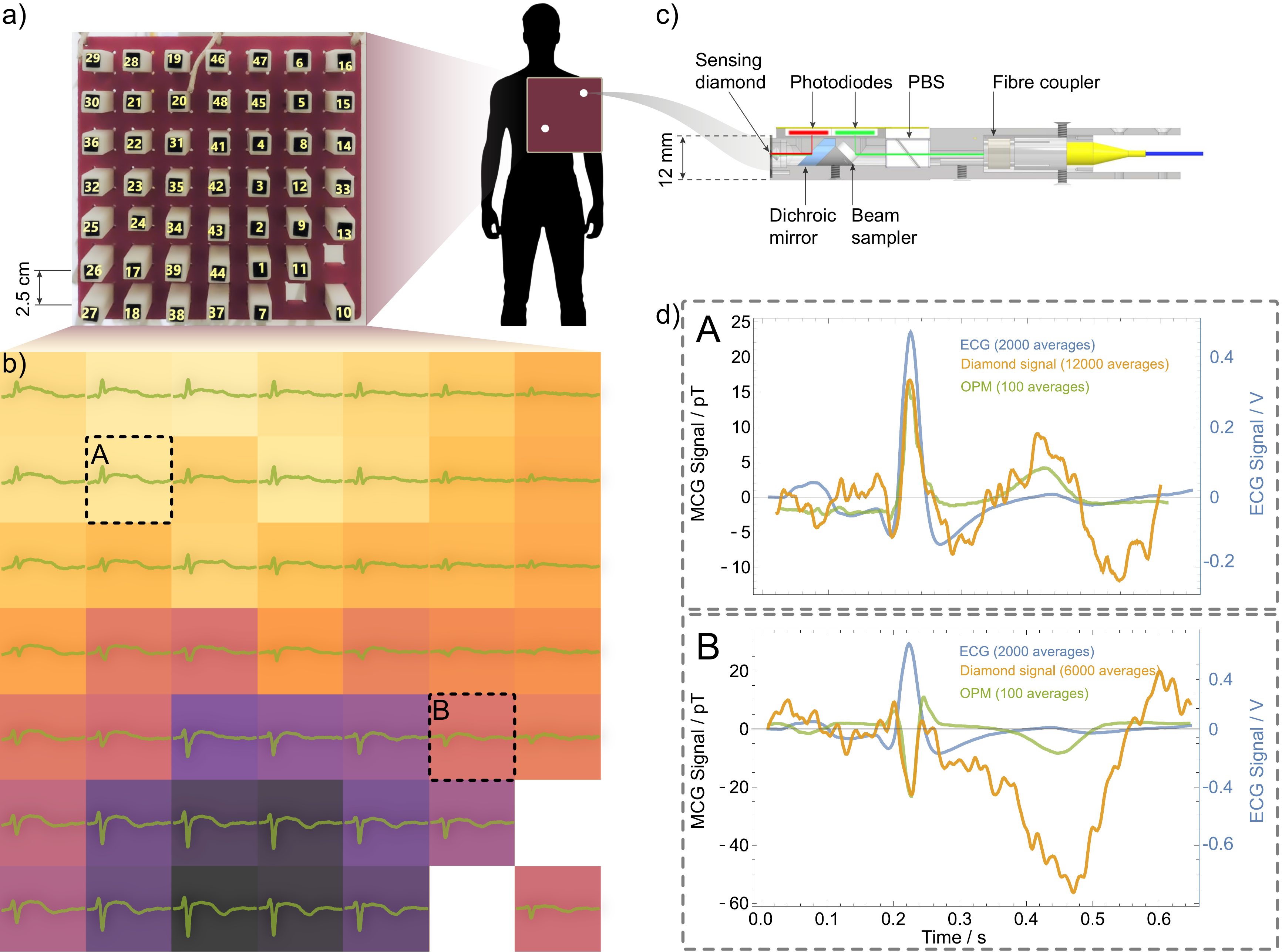}
    \caption{a) OPM array with sensor labeling and its position on the subject's chest. The position of the diamond sensor measurements are indicated as the white dots on the torso. b) The 47 different OPM MCG signals after band-pass filtering and averaging 100 times. The color overlay indicates the amplitude of the QRS complex. A and B mark the positions of the diamond sensor measurements. c) Schematic of the JGU sensor [photographs shown in Fig.\ref{fig:1} a)] with indication of the main components. Working principle and more details can be found in \cite{zulfpaper}. PBS: Polarizing beam splitter. d) The resulting OPM, ECG, and diamond signal traces for MCG signal detection at the positions of A and B with respect to the OPM array.}
    \label{fig:2}
\end{figure}

To inform the diamond sensor positions and confirm the signal sizes, the MCG signal over the chest area of the subject was mapped using a FieldLine OPM array  in a magnetically shielded room (2-layer $\mu$-metal shielding) at the Fraunhofer Institute for Physical Measurement Techniques in Freiburg (IPM). The array-sensor positions [see Fig.\,\ref{fig:2}\,a)] were indicated on form-fitting shirts to identify the positions for the diamond sensor measurement conducted afterwards. The sensor spacing in the array is 2.5\,cm and the stand-off distance to the chest was at least 5\,cm for all sensors. The array's MCG signals were visible in real-time. The OPM-array data were acquired during an approximately 5\,min period. The OPM-array MCG traces in Fig.\,\ref{fig:2}\,b) show the data from all sensors averaged over 100 heartbeats. Before averaging the OPM signals, a digital band-pass filter (pass band between 3 and 30\,Hz) was applied to the data. The MCG peak was detected and used to align individual time traces for averaging. The OPM array traces are shown in relation to the chest in Fig.\,\ref{fig:2}\,b). The diamond sensor MCG arrangement is shown in Fig.\,\ref{fig:2}\,c). The MCG was acquired at positions A and B and averaged over 12,000 and 6,000 heartbeats, respectively. It was compared to the OPM data at the respective positions and the ECG trace in Fig.\,\ref{fig:2}\,d). The measured MCG peak amplitudes between OPM and diamond sensor were comparable and consistent with values in the literature\,\cite{Heliyon2024Su,Zheng2020OPG,YoungJin2019APL}.

Despite the differences in offset distance, sensitive volume, and sensitivity axis orientation of the respective sensors to the chest, the signals of both sensors coincidentally had matching QRS complex amplitudes but differed significantly in the T-wave shape. This can be caused by positioning and sensing volume differences. The OPM array's MCG signatures for each position show a large variability in the T-wave signal around the measured position consistent with literature (e.g. \cite{mcgsquid}). So position variation lead to a variation in T-wave size and shape. Furthermore, the difference in sensitive volumes of the diamond and OPM sensors also affect the recorded T-wave morphology due to signal differences and directional variation over the sensing volume, and the distance to the source.

\subsection{Gradiometry performance}
Gradiometric configurations mitigate a key challenge in biomagnetic sensing: strong and spatially uniform environmental noise, which in clinical or laboratory environments can exceed the amplitude of cardiac signals by several orders of magnitude. By measuring the field difference between two spatially separated sensors, a gradiometer extracts the local field gradient. 
This differential measurement suppresses common-mode magnetic noise while preserving spatially localized biomagnetic signals. In practice, a well-matched gradiometer can achieve common-mode rejection ratios (CMRRs) exceeding 60\,dB, equivalent to a suppression factor of 1000 for homogeneous noise fields\,\cite{Taulu2014NovelNoiseReduction}.

\begin{figure}
    \centering
    \includegraphics[width=1\columnwidth]{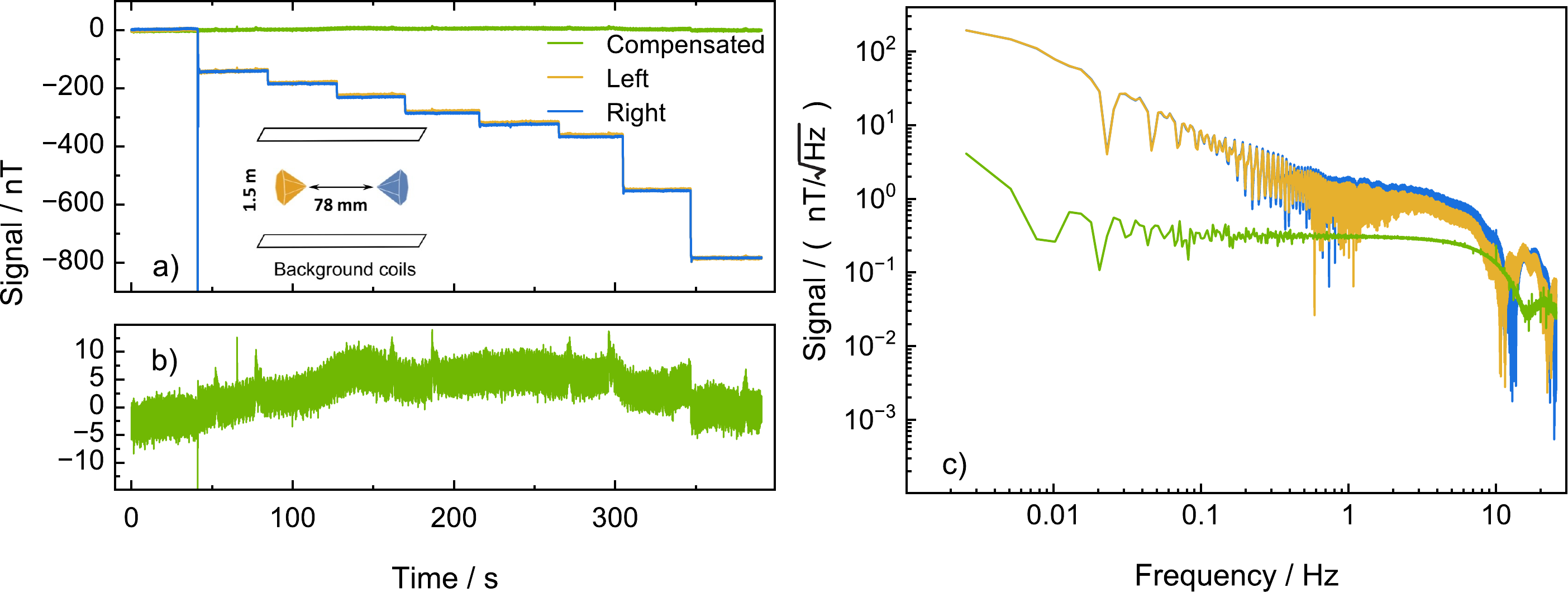}
    \caption{a) Time-domain traces from the two diamond sensors of ZAQuant used to characterize gradiometry performance, (“right” sensor in blue, “left” sensor in orange) together with their differential signal (green), recorded under different background-coil magnetic fields. Inset: Schematic illustrating the two diamond sensors, used to assess the gradiometric performance. b) Zoomed in differential signal time trace.
    c) Amplitude spectral density corresponding to the three magnetic traces shown in panel (a).}
    \label{fig:3}
\end{figure}

At ZAQuant, two NV sensor heads separated by 78\,mm were placed within background-field Helmholtz coils ($\approx$ 1.5\,m diameter) inside the single-layer magnetically shielded room (Fig.\,\ref{fig:3}a, inset). Controlled background fields were applied to simulate ambient interference, while localized signals were introduced to emulate biomagnetic sources. 

Magnetic test fields on the order of several hundred nanotesla were applied, and the resulting time-domain traces from both sensors were recorded. As shown in Fig.\,\ref{fig:3}\,a), a simple subtraction of the two sensor signals was sufficient to cancel the applied test field and to substantially reduce the overall noise. The first 60\,s data of Fig.\,\ref{fig:3}\,a) of the two magnetometers is displayed against each other in Fig.\,\ref{fig:3}\,c) showing weakly correlated noise components. The performance of the gradiometer scheme can be seen in Fig.\,\ref{fig:3}\,d). It shows the noise spectral density [derived from the data in  Fig.\,\ref{fig:3}\,a)] of both sensors and of the subtracted signal. The noise density of the individual sensors was larger than 100\,nT$/\sqrt{\mathrm{Hz}}$ at low mHz-frequencies and was suppressed by almost two orders of magnitude in the gradiometry trace. 
The linear range of the gradiometer is ultimately limited by the magnetic resonance linewidth (ZAQuant sensor: $\sim$200\,kHz), degrading gradiometric suppression beyond $\sim\pm$500\,nT. This range can be extended via active frequency tracking and closed-loop feedback, enabling real-time operation under field variations of up to several microtesla, relevant for dynamic environments present for example in a neurosurgery theater. Furthermore, the scalability of the diamond platform supports gradiometric baselines from millimeter to meter scale, allowing system-level optimization for specific biomedical tasks.
These results demonstrate a clear pathway for implementing diamond-sensor MCG detection in realistic, noisy environments using a gradiometric configuration.

\section{Discussion and outlook}
\label{Discussion}

This work demonstrates human MCG with NV-based sensors, marking a crucial step towards clinical translation of quantum technologies. However, a significant performance gap remains between current capabilities and the requirements for routine clinical or industrial use, primarily due to sensitivity limitations.
The following questions need to be addressed to overcome these challenges: 
How can sensor sensitivity be increased to allow real-time detection with diagnostically sufficient signal-to-noise ratios, and, what approaches exist to enhance biomagnetic signal amplitudes at the sensor position?
Furthermore, we discuss broader applications and a potential pathway towards large scale deployment.

\begin{figure}
    \centering
    \includegraphics[width=0.6\columnwidth]{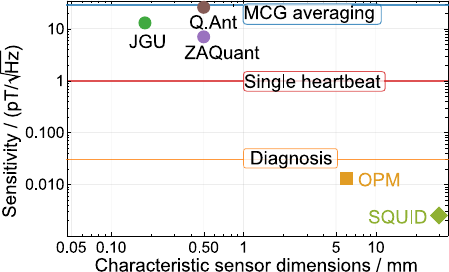}
    \caption{Comparison of sensors used for MCG detection, including the present work, characterized by their sensitivity and a representative linear dimension of the sensing volume. Green: JGU, purple: ZAQuant, brown: Q.ANT, SQUID\,\cite{Matlashov2011}, OPM\,\cite{Alem2023OPMMEG}}
    \label{fig:4}
\end{figure}
\subsection{Towards live detection of biomagnetic signals: Sensitivity and signal enhancement prospects}
Recent NV-based magnetometry systems have achieved sub-10\,pT/$\sqrt{\mathrm{Hz}}$ sensitivities\,\cite{barryensing} positioning them as promising candidates within the broader magnetometry landscape. However, for clinical applications such as MCG and magnetoencephalography (MEG), sensitivities in the sub-pT/$\sqrt{\mathrm{Hz}}$ range—currently reached with SQUIDs and OPMs—remain necessary\,\cite{Alem2023OPMMEG,mcgsquid}. An illustration of the achieved sensitivities within this work and estimated sensitivity targets for clinical relevance are displayed along with different sensing modalities and their characteristic sensor dimensions in Fig.\,\ref{fig:4}. The performance of NV-based sensors allows the observation of averaged MCG but clinical relevance requires larger signal-to-noise\,\cite{SWAIN2020101664}.

The baseline for live detection in Fig.\,\ref{fig:4} is determined by the measured signal amplitudes. With sensitivities on the order of 10\,pT$/\sqrt{\mathrm{Hz}}$, a bandwidth of approximately 30\,Hz, and a single-shot root-mean-square (RMS) noise floor of 50\,pT, reliable detection of the average 25\,pT R-peaks in the MCG signals with a signal-to-noise ratio of 5 would require a sensitivity of about 1\,pT$/\sqrt{\mathrm{Hz}}$. This corresponds to roughly an order-of-magnitude improvement over the sensors used in this work, but such performance has already been achieved in other NV-center setups\,\cite{barryensing}.
Bridging this gap necessitates identifying and addressing the dominant limitations of current NV implementations.
The NV ensemble magnetometers discussed in this work operate using continuous-wave (CW) protocols, which are approaching their practical sensitivity limits\,\cite{2019BarryNVReview}. In these systems, a dominant noise contribution arises from photon shot noise, scaling as $1/\sqrt{P}$ with $P$ being the detected fluorescence power. While increasing optical excitation can reduce this noise, further improvements are constrained by power broadening, sample heating, and laser-induced decoherence. 

To surpass this limit, several quantum-enhanced sensing strategies have been proposed and experimentally explored.  
\textit{Infrared absorption magnetometry} offers a pathway for high photon collection by spectroscopically interrogating the singlet transition of the NV center in diamond via the absorption of an infrared laser, potentially lowering the photon shot noise limited sensitivity further\,\cite{PhysRevApplied.8.044019,PhysRevApplied.23.054019}.
\textit{Pulsed magnetometry} schemes exploit NV spin coherence more efficiently and suppress low-frequency technical noise via dynamical decoupling sequences, with demonstrated sensitivities reaching 490\,fT/$\sqrt{\mathrm{Hz}}$\,\cite{barryensing}. \textit{Microwave-cavity-enhanced detection}, by coupling NV ensembles to microwave (MW) resonators under strong coupling conditions, may enable an order-of-magnitude sensitivity improvement\,\cite{eisenach2021cavity,wang2025exceptional}. \textit{Many-body quantum effects}, including spin squeezing and quantum gain amplification~\cite{gao2025signal,wu2025spin}, offer potential routes to surpass the standard quantum limit, although their practical implementation remains at an early stage.

While quantum-enhanced strategies hold promise for long-term sensitivity improvements, their experimental complexity currently limits their near-term applicability. Magnetic flux concentrators offer a practical alternative for immediate performance enhancement \cite{fescenko2020diamond,xie2021hybrid}. By channeling magnetic flux from a larger volume into the sensing region, these passive structures can amplify the local field experienced by the NV ensemble, yielding more than two orders of magnitude signal enhancement without requiring changes to the underlying measurement protocol\,\cite{fescenko2020diamond,keigomcg}.
In addition to their technical simplicity, flux concentrators can be matched to many biomedical applications. An associated reduction in spatial resolution---due to flux collection over larger areas---can be acceptable in several biomagnetic sensing scenarios such as cardiac monitoring, where centimeter-scale localization is sufficient. 



\subsection{Application opportunities beyond cardiac sensing and pathways towards widespread adoption}

The unique combination of high dynamic range, gradiometric noise rejection, and scalable geometry positions NV magnetometers for several emerging biomedical applications. In surgical oncology, NV-based gradiometry enables sentinel lymph node localization through magnetic nanoparticle tracer detection\,\cite{newman2025endoscopic} and also supports intra-operative nerve monitoring in unshielded environments\,\cite{Azargoshasb2022}.
The spatial gradient sensitivity facilitates non-invasive separation of maternal and fetal cardiac signals for prenatal monitoring\,\cite{Strasburger2008Magnetocardiography}.
Most significantly, room-temperature NV gradiometers could enable portable MEG systems, enabling new avenues for neurological diagnostics or next-generation brain-computer interfaces (BCIs). While current MEG sensitivity requirements remain stringent ($<$10\,fT/$\sqrt{\mathrm{Hz}}$), magnetic flux concentrators ($>$100$\times$ gain) combined with gradiometric suppression ($>$60\,dB) could bridge this gap. The planar NV geometry enables dense arrays with millimeter spacing, potentially exceeding conventional SQUID-based spatial resolution. Initial applications may focus on high-amplitude signals such as auditory evoked responses or epileptic discharges (100--500\,fT)\,\cite{KOWALCZYK2021117497}.

From an industrial perspective, the large-scale deployment of diamond-based quantum sensors requires progress along three key axes. First, a commercially relevant use case must be identified in which the technology offers clear advantages. Cardiovascular diseases, for example, represent a major public health concern, with an estimated economic burden of \euro\,46.4\,billion annually
in Germany\,\cite{DestatisCostOfIllness}. Improved diagnostic throughput and continuous, contactless monitoring protocols may significantly reduce treatment costs, making this an attractive application area.

Second, a proof-of-concept demonstration addressing the target use case is essential prior to high-cost product development. The present work provides initial evidence and a clear path towards NV-diamond magnetometers serving as an advanced diagnostic platform for cardiac monitoring, potentially surpassing conventional ECG methods in sensitivity and spatial resolution.

The operational complexity of traditional magnetic sensors has long prevented MCG from challenging the ubiquity of the ECG. Our current findings are a first step highlighting how diamond magnetometry may fundamentally shift this paradigm. By combining the robustness of solid-state sensors with the full vector information content of magnetic fields, this technology paves the way towards clinical translation. Ultimately, diamond sensors hold the promise of non-invasive, high-resolution source localization, offering a new dimension in cardiac diagnostics beyond current capabilities, that may also extent to other applications areas such as non-invasive measurement of neural activity for next-generation brain interfacing.

Non-invasive source localization represents an unmet need in contemporary arrhythmia diagnostics. Current clinical workflows often require invasive electro-anatomical mapping to precisely identify arrhythmogenic foci, particularly in patients with atypical or intermittent rhythm disturbances. High spatial-density magnetocardiographic recordings offer a potential non-invasive alternative by enabling spatial mapping of cardiac activation patterns, which could guide patient selection, procedural planning, and hypothesis generation prior to invasive electrophysiological intervention.

Finally, commercialization requires miniaturization, user-friendly interfaces, and software-driven data acquisition and interpretation. Ideally, such systems translate raw sensor output into clinically relevant diagnostic information. With further improvements in sensitivity and array integration, diamond-based quantum sensors enable real-time, clinically actionable biomagnetic measurements.

\begin{table}[h]
\caption{Comparison of electrocardiography (ECG) and magnetocardiography (MCG) technologies with different sensing modalities}\label{tab1}
\centering
\begin{tabular}{@{}p{2.3cm}p{4cm}p{2cm}p{2cm}p{2cm}@{}}
\toprule
& \multicolumn{1}{@{}c@{}}{ECG (Electrocardiography)} 
& \multicolumn{3}{>{\centering\arraybackslash}p{7cm}}{MCG (Magnetocardiography)} \\
\cmidrule{2-2}\cmidrule{3-5}

Quantum sensor &   &
\multicolumn{1}{>{\centering\arraybackslash}p{2cm}}{SQUID} &
\multicolumn{1}{>{\centering\arraybackslash}p{2cm}}{OPM} &
\multicolumn{1}{>{\centering\arraybackslash}p{2cm}}{Diamond} \\
\midrule

Principle & Measures electric potentials 
& \multicolumn{3}{>{\centering\arraybackslash}p{7cm}}{Measure magnetic fields} \\
\midrule

Signal  
& Electric field, strongly affected by tissue conductivity (lungs, fat, fluids distort signals) 
& \multicolumn{3}{>{\arraybackslash}p{7cm}}{Magnetic field, hardly attenuated by tissues, minimally affected by conductivity distribution} \\
\midrule

\multirow[t]{2}{2.3cm}{Contact method} &
\multirow[c]{2}{4cm}{\centering Attached to the body surface\par} &
\multicolumn{3}{>{\centering\arraybackslash}p{7cm}}{Non-contact} \\
\cmidrule(lr){3-5}
& & Safety distance due to cryostat & Safety distance due to cell temperature & Skin contact \\
\midrule

\multirow[t]{2}{2.3cm}{Spatial resolution } &
\multirow[c]{2}{4cm}{\centering cm level\par} &
\multicolumn{3}{>{\centering\arraybackslash}p{7cm}}{High, enabling precise localization and imaging} \\
\cmidrule(lr){3-5}
& & mm level  & mm level  & µm level \\
\midrule

Advantages  
& - Mature, inexpensive 
& \multicolumn{3}{>{\arraybackslash}p{6cm}}{
- Non-contact, high patient comfort 
} \\

&- Easy to operate, widely available 
& \multicolumn{3}{>{\arraybackslash}p{7cm}}{
- unaffected by tissue conductivity 
} \\

&- Suitable for dynamic monitoring (Longterm-, exercise ECG)
& \multicolumn{3}{>{\arraybackslash}p{7cm}}{
- Detects subtle or deep abnormalities, useful for arrhythmia source localization and early ischemia detection, access to fetal MCG }
\\

\midrule

Disadvantages 
& - Strongly influenced by tissue properties
& \multicolumn{3}{>{\arraybackslash}p{7cm}}{
- Expensive 
} \\

& - Limited localization ability
& \multicolumn{3}{>{\arraybackslash}p{7cm}}{
- Significant environmental constraints requiring a shielded room or  gradiometric schemes 
}\\

&- Low sensitivity to early or subtle abnormalities
& \multicolumn{3}{>{\arraybackslash}p{7cm}}{
- Not yet widely adopted in clinical practice
} \\

\midrule

Application
& - Arrhythmia
& \multicolumn{3}{>{\arraybackslash}p{7cm}}{
- Arrhythmia origin localization 
} \\

scenarios&- Myocardial infarction 
& \multicolumn{3}{>{\arraybackslash}p{7cm}}{
- Detection of micro-ischemia 
} \\

&- Routine clinical examinations
& \multicolumn{3}{>{\arraybackslash}p{7cm}}{
- Electronic pathway reconstruction 
} \\
&
& \multicolumn{3}{>{\arraybackslash}p{7cm}}{
- Fetal MCG
} \\

\midrule

Popularity 
& Established as the clinical “gold standard”
& \multicolumn{3}{>{\arraybackslash}p{7cm}}{Mainly used in research and specialized clinical centers
} \\

\botrule
\end{tabular}
\end{table}

\section{Methods}

In this work, we developed three NV-diamond-based magnetometers.

At JGU, a zero-field magnetometry protocol was employed, achieving sensitivities comparable to bias-field optically detected magnetic resonance (ODMR) spectroscopy\,\cite{substractionodmr}. The zero-background compatibility enabled a direct comparison between diamond- and OPM-based MCG detection. The protocol relied on ODMR spectroscopy at zero magnetic bias field. A modulated magnetic field (6.12 kHz, µT amplitude) was applied along the [100] crystallographic direction of the diamond, allowing to retrieve a magnetically sensitive signal from NV centers along all crystal axes without requiring a static bias field\,\cite{zerohuijie,zulfpaper,lenz2021magnetic}. 

The central microwave frequency at 2.87\,GHz was mixed with a 2.16\,MHz radio frequency signal to address all hyperfine level, enhancing resonance contrast and improving sensitivity. The central microwave frequency was modulated (modulation frequency 140\,kHz, modulation amplitude 100\,kHz) and locked on resonance using the demodulated dispersive signal as the error signal in a feedback loop. The MW signal was delivered to the sensing diamond using a custom made printed circuit board (PCB). At zero bias field, this microwave frequency locking renders the magnetically sensitive signal largely insensitive to temperature variations, enhancing the stability of long-term measurements.

The design of the sensing diamond and the diamond lens, optimized for photoluminescence collection, has been described in detail in reference\,\cite{diamondoptics}. The sensing diamond and lens are mounted on a diamond plate for thermal management, which is in turn mounted on an aluminum housing (sensor head diameter: 12\,mm). The housing integrates optics for laser delivery from a high-power fiber, collection of photoluminescence through a dichroic mirror, and detection via a differential photodiode board. The differential detection scheme minimizes the influence of laser noise on the signal. A customized holder for the optical fiber output coupler is also integrated into the sensor head. The sensor design is described in further detail in reference\,\cite{zulfpaper}.

Fiber and signal cables are routed to a rack housing a breadboard, which supports a Coherent Verdi G5 laser head with integrated optics, a Zurich Instruments lock-in amplifier for data processing, associated microwave electronics, and a personal computer (PC) for instrumentation control and data analysis.

The Q.ANT sensor employed a similar sensor head architecture, maintaining a compact overall footprint with a sensor head volume of 1\,$\mathrm{dm}^3$. A fiber-coupled OBIS laser from Coherent provided 100\,mW of power at the fiber output was used to pump the sensing diamond. The sensor system integrates an field-programmable gate array (FPGA) for control, microwave generation, and a diamond (1\,mm × 1\,mm × 500\,µm) with a truncated pyramid geometry containing approximately 1.7\,ppm NV\textsuperscript{–}, as determined by electron paramagnetic resonance (EPR) spectroscopy.

Fluorescence from the diamond was acquired using a low-noise balanced photodiode board with a differential amplifier, which simultaneously monitored part of the laser light to suppress intensity fluctuations, similar to the approach used at JGU. To resolve all four NV-resonance pairs corresponding to the four NV orientations in the diamond lattice, a bias field of approximately 1\,mT was applied via a in-house fabricated Halbach ring.

The sensor interfaces with a control PC via universal serial bus connection (USB-C), running software that includes a graphical user interface for sensor control and real-time data visualization. A sophisticated miniaturized optical design collected diamond fluorescence with an estimated efficiency of roughly 40\,\%, as determined from optical simulations.

At ZAQuant, the diamond sensors were similarly designed as the Q.ant sensors.
As shown in Fig.\,\ref{fig:3}(b), the optical module consisted of a green laser excitation path and a compact fluorescence detection stack. The laser was directed through a dichroic cube and guided by a glass rod toward a pyramidal diamond mounted on a compound parabolic concentrator (CPC), which enhanced fluorescence collection. The emitted fluorescence was reflected by the dichroic cube, spectrally filtered with a \SI{695}{nm} long-pass filter, and detected using a balanced photodetection scheme to suppress laser intensity noise. A static magnetic field of approximately \SI{1}{mT} was applied by a permanent magnet aligned with one of the NV axes; for gradiometric operations, two identical sensor heads were arranged in a mirrored geometry, such that the sensors face each other (see Fig.\,\ref{fig:3} a) represented as symbolically as colored diamonds in the inset). This allowed precise alignment of the magnetometers.
The electronics subsystem included microwave signal generation, and signal readout with a Zurich Instruments HF2LI lock-in amplifier. A double-resonance scheme combined with hyperfine driving was implemented to address all three hyperfine transitions of a selected NV axis similar to the JGU sensor. A 3\,kHz MW amplitude modulation was applied to enable lock-in detection with enhanced sensitivity. This modulation frequency balances signal contrast and noise suppression\,\cite{subpicowracht}. The same architecture was duplicated for the second spin transition with opposite modulation phase, and the resulting microwave signals were amplified and delivered to the diamond via a loop antenna.

\begin{table}[ht]
\begin{tabular}{|c|c|c|c|}
\toprule
\textbf{Property} 
& \textbf{JGU} 
& \textbf{ZAQuant} 
& \textbf{Q.Ant} \\
\hline
 Diamond dimension (w x l x h) 
&  $0.5 \times 0.5 \times 0.18~\mathrm{mm}^3$ 
& $0.5\,\times 0.5\times 0.5\, \mathrm{mm}^3$ 
& $1\,\times 1\times 0.5\, \mathrm{mm}^3$\\ 
\hline

Sensitivity 
& $13~\mathrm{pT}/\sqrt{\mathrm{Hz}}$ 
& $7~\mathrm{pT}/\sqrt{\mathrm{Hz}}$ 
& $26~\mathrm{pT}/\sqrt{\mathrm{Hz}}$ \\ \hline

Magnetic bias
& Zero DC bias
& 1 mT bias    
& 1 mT bias \\ 
\hline  

Modulation 
&  magnetic mod.
&  MW amplitude mod.   
&  MW amplitude mod. \\ 
\hline  
Operating location 
& multi-layer shielded   
& single layer shielded  
& unshielded \\

\hline 
\end{tabular}
\caption{Overview of the three diamond sensors.} 
\label{tab:sensor_comparison}
\end{table}

 \section{Acknowledgments}
 This work was supported by the EU, project HEU-RIA-MUQUABIS-101070546, by the DFG, project FKZ: SFB 1552/1 465145163 and by the German Federal Ministry
of Research, Technology and Space (BMFTR) within the Quantumtechnologien program via the DIAQNOS project (project no. 13N16455). In addition it has been supported by the by the Ministerium für Wirtschaft, Arbeit und Tourismus (InvestBW), project VitalQ,
 BW$8_1166_2170$, by the Helmholtz Association project Quantum Sensing for Fundamental Physics
(QS4Physics) from the Innovation pool of the research
field Helmholtz Matte and by the BMFTR within the clusters for future program, project QHMI2 - 03ZU2110FA. We thank Hamamatsu for providing the non-magnetic photodiodes used in the JGU sensor as samples.

\section{Data availability}
The noise characterization and MCG analysis data are available upon reasonable request from the authors. 

\section{Code availability}
The code used in manuscript is available from the corresponding authors upon request.

\section{Ethics Declaration}
No Ethics approval was required for the non-invasive self-experiments (personal communication of the local Ethics Committee of the University of Freiburg to TB).

\subsection*{Competing interests}

The authors declare no competing interests.

\bibliography{sn-bibliography}

@article{Oogane2021SubpT,
  author       = {Mikihiko Oogane and Kosuke Fujiwara and Akitake Kanno and Takafumi Nakano and Hiroshi Wagatsuma and Tadashi Arimoto and Shigemi Mizukami and Seiji Kumagai and Hitoshi Matsuzaki and Nobukazu Nakasato and Yasuo Ando},
  title        = {Sub-pT magnetic field detection by tunnel magneto-resistive sensors},
  journal      = {Applied Physics Express},
  volume       = {14},
  number       = {12},
  pages        = {123002},
  year         = {2021},
  publisher    = {IOP Publishing},
  doi          = {10.35848/1882-0786/ac3809},
  url          = {https://iopscience.iop.org/article/10.35848/1882-0786/ac3809}
}

@article{Karo2016First36Channel,
  author       = {Hikaru Karo and K. Shimoda and Yoshiaki Maeda and Ichiro Sasada},
  title        = {The first 36 channel fluxgate-sensor-array for the MCG measurement},
  journal      = {IEEJ Transactions on Sensors and Micromachines},
  volume       = {136},
  number       = {6},
  pages        = {224--228},
  year         = {2016},
  doi          = {10.1541/ieejsmas.136.224},
  url          = {https://doi.org/10.1541/ieejsmas.136.224},
  publisher    = {The Institute of Electrical Engineers of Japan}
}

@article{Azargoshasb2022,
  title        = {Advancing intraoperative magnetic tracing using 3D freehand magnetic particle imaging},
  author       = {Azargoshasb, Samaneh and Molenaar, Lennert and Rosiello, Giuseppe and Buckle, Tessa and van Willigen, Danny M. and van de Loosdrecht, Melissa M. and Welling, Mick M. and Alic, Lejla and van Leeuwen, Fijs W.~B. and Winter, Alexander and van Oosterom, Matthias N.},
  journal      = {International Journal of Computer Assisted Radiology and Surgery},
  volume       = {17},
  number       = {1},
  pages        = {211--218},
  year         = {2022},
  doi          = {10.1007/s11548-021-02458-2},
  pmid         = {34333740}
}

@article{KOWALCZYK2021117497,
title = {Detection of human auditory evoked brain signals with a resilient nonlinear optically pumped magnetometer},
journal = {NeuroImage},
volume = {226},
pages = {117497},
year = {2021},
issn = {1053-8119},
doi = {https://doi.org/10.1016/j.neuroimage.2020.117497},
url = {https://www.sciencedirect.com/science/article/pii/S1053811920309824},
author = {Anna U. Kowalczyk and Yulia Bezsudnova and Ole Jensen and Giovanni Barontini}
}

@article{Alem2023OPMMEG,
  title        = {An integrated full-head OPM-MEG system based on 128 zero-field sensors},
  author       = {Alem, Orang and Hughes, K. Jeramy and Buard, Isabelle and Cheung, Teresa P and Maydew, Tyler and Griesshammer, Andreas and Holloway, Kendall and Park, Aaron and Lechuga, Vanessa and Coolidge, Collin and Gerginov, Marja and Quigg, Erik and Seames, Alexander and Kronberg, Eugene and Teale, Peter and Knappe, Svenja},
  journal      = {Frontiers in Neuroscience},
  volume       = {17},
  pages        = {1190310},
  year         = {2023},
  publisher    = {Frontiers Media SA},
  doi          = {10.3389/fnins.2023.1190310},
  url          = {https://www.frontiersin.org/articles/10.3389/fnins.2023.1190310/full},
  pmcid        = {PMC10303922},
  pmid         = {37389367}
}

@article{Matlashov2011,
  author       = {Andrei N. Matlashov and Larry J. Schultz and Michelle A. Espy and Robert H. Kraus and Igor M. Savukov and Petr L. Volegov and Caroline J. Wurden},
  title        = {SQUIDs vs. Induction Coils for Ultra-Low Field Nuclear Magnetic Resonance: Experimental and Simulation Comparison},
  journal      = {IEEE Transactions on Applied Superconductivity},
  year         = {2011},
  volume       = {21},
  number       = {3},
  pages        = {465--468},
  doi          = {10.1109/TASC.2010.2089402},
  pmcid        = {PMC3131692},
  note         = {PMID: 21747638}
}

@article{barryensing,
author = {John F. Barry and Matthew H. Steinecker and Scott T. Alsid , Jonah Majumder and Linh M. Pham and Michael F. O’Keeffe and and Danielle A. Braje},
title = {Sensitive ac and dc magnetometry with nitrogen-vacancy-center ensembles in diamond
},
journal = {Phys. Rev. Applied 22, 044069, 2024},
doi= {https://doi.org/10.1103/PhysRevApplied.22.044069},
}

@article{zulfpaper,
  title = {Zero- to low-field J-spectroscopy with a diamond magnetometer},
  author = {M. Omar and J. Xu and R. Kircher and P. Sharbati and S. Zhang and G. Chatzidrosos and J. Eills and R. Picazo-Frutos and
D. Budker and D. A. Barskiy and A. Wickenbrock},
  year = {2026},
  journal = {arxiv preprint https://arxiv.org/abs/2512.05776 },
}

@article{fft,
  author = {G. Heinzel and A. Rüdiger and R. Schilling},
  title = {Spectrum and Spectral Density Estimation by the Discrete Fourier Transform (DFT), Including a Comprehensive List of Window Functions and Some New At-Top Windows},
  journal = {Max-Planck-Institut für Gravitationsphysik Report},
  year = {2002},
  url = {https://holometer.fnal.gov/GH_FFT.pdf}
}

@article{strainimaging,
author = {S. Knauer and J. P. Hadden and J. G. Rarity },
title = {In-situ measurements of fabrication induced strain in diamond photonic-structures using intrinsic colour centres

},
journal = {npj Quantum Information volume 6, Article number, 50, 2020},

doi = {10.1038/s41534-020-0277-1},
url = {https://www.nature.com/articles/s41534-020-0277-1},

}

@article{temperature2,
author = {J. H. Shim and S.-J. Lee and S. Ghimire and J. I. Hwang and K.-G. Lee and K. Kim and M. J. Turner and C. A. Hart and R. L. Walsworth and S. Oh },
title = {Multiplexed sensing of magnetic field and temperature in real time using a nitrogen
vacancy spin ensemble in diamond
},
journal = {Phys. Rev. Applied 17, 014009 , 2022},

doi = { https://doi.org/10.1103/PhysRevApplied.17.014009},
url = {https://journals.aps.org/prapplied/abstract/10.1103/PhysRevApplied.17.014009},

}

@article{electric1,
author = {Dolde, F. and Fedder, H. and Doherty, M. },
title = { Electric-field sensing using single diamond spins},
journal = { Nature Phys 7, 459–463 ,(2011)},

doi = { https://doi.org/10.1038/nphys1969
},
url = {https://www.nature.com/articles/nphys1969},

}

@article{reviewmcg,
  title = {Diagnostic value of magnetocardiography in coronary artery disease and cardiac arrhythmias: A review of clinical data
},
  author = {Joey S.W. Kwonga and  Boris Leithäuserb and Jai-Wun Parkb Cheuk-Man Yu},
  journal = {International Journal of Cardiology},
  volume = {167},
  issue = {5},
  pages = {1835-1842},
  numpages = {4},
  year = {2013},
  month = {September},
  doi = {10.1016/j.ijcard.2012.12.056},
  url = {https://www.sciencedirect.com/science/article/pii/S0167527312016853}
}

@article{diamondoptics,
  title = {Diamond-optic enhanced photon collection efficiency for sensing with nitrogen-vacancy centers
},
  author = { Omar,Muhib and  Conta,Andreas and  Westerhoff,Andreas and  Hasse,Raphael and  Chatzidrosos,Georgios and  Budker,Dmitry and  Wickenbrock,Arne

},
  journal = {Optics Letters},
  volume = {48},
  issue = {10},
  pages = { 2512-2514 },
  numpages = {},
  year = {2023},
  month = {},
  publisher = {},
  doi = { https://doi.org/10.1364/OL.486998},
  url = {https://opg.optica.org/ol/abstract.cfm?uri=ol-48-10-2512}
}

@article{fluxmcg,
  title = {Noninvasive magnetocardiography of a living rat based on a diamond quantum sensor
},
  author = { Yu,Ziyun and  Xie,Yijin and  Jin,Guodong and  Zhu,Yunbin and  Zhang,Qi and  Shi,Fazhan
and  Wan,Fang-yan and  Luo,Hongmei and Tang,Ai-hui and
 Rong,Xing},
  journal = {Phys. Rev. Applied},
  volume = {21},
  issue = {064028},
  pages = {},
  numpages = {},
  year = {2024},
  month = {},
  publisher = {},
  doi = { https://doi.org/10.1103/PhysRevApplied.21.064028},
  url = {https://journals.aps.org/prapplied/abstract/10.1103/PhysRevApplied.21.064028}
}

@article{keigomcg,
  title = {Millimetre-scale magnetocardiography of living rats with thoracotomy
},
  author = { Arai,Keigo and  Kuwahata,Akihiro and  Nishitani,Daisuke and  Fujisaki,Ikuya and  Matsuki, Ryoma and  Nishio,Yuki and  Xin,Zonghao and  Cao,Xinyu amd  Hatano,Yuji and  Onoda,Shinobu and   Shinei, Chikara and Miyakawa,Masashi and  Taniguchi,Takashi and  Yamazaki,Masatoshi and  Teraji,Tokuyuki and  Ohshima,Takeshi and  Hatano, Mutsuko and  Sekino,Masaki and  Iwasaki,Takayuki },
  journal = {Communications Physics},
  volume = {5},
  issue = {200},
  pages = {},
  numpages = {},
  year = {2022},
  month = {},
  publisher = {},
  doi = {https://doi.org/10.1038/s42005-022-00978-0},
  url = {https://www.nature.com/articles/s42005-022-00978-0}
}

@article{mcgsquid,
  title = {Multi-channel Magnetocardiogardiography System Based on Low-Tc SQUIDs in an Unshielded Environment },
  author = { Konga,Xiangyan and  Zhanga,Shulin and  Wanga,Yongliang and  Zenga,Jia and  Xie,Xiaoming},
  journal = {Physics Procedia },
  volume = {36},
  issue = {},
  pages = {286 – 292},
  numpages = {6},
  year = {2012},
  month = {},
  publisher = {Elsevier B.V},
  doi = {doi.org/10.1016/j.phpro.2012.06.161},
  url = {https://www.sciencedirect.com/science/article/pii/S1875389212020986?ref=pdf_download&fr=RR-2&rr=90bc96d0c8c73a5a}
}

@article{PhysRevApplied.23.054019,
  title = {Towards high-sensitivity magnetometry with nitrogen-vacancy centers in diamond using the singlet infrared absorption},
  author = {Tayefeh Younesi, Ali and Omar, Muhib and Wickenbrock, Arne and Budker, Dmitry and Ulbricht, Ronald},
  journal = {Phys. Rev. Appl.},
  volume = {23},
  issue = {5},
  pages = {054019},
  numpages = {12},
  year = {2025},
  month = {May},
  publisher = {American Physical Society},
  doi = {10.1103/PhysRevApplied.23.054019},
  url = {https://link.aps.org/doi/10.1103/PhysRevApplied.23.054019}
}

@article{PhysRevApplied.8.044019,
  title = {Miniature Cavity-Enhanced Diamond Magnetometer},
  author = {Chatzidrosos, Georgios and Wickenbrock, Arne and Bougas, Lykourgos and Leefer, Nathan and Wu, Teng and Jensen, Kasper and Dumeige, Yannick and Budker, Dmitry},
  journal = {Phys. Rev. Appl.},
  volume = {8},
  issue = {4},
  pages = {044019},
  numpages = {5},
  year = {2017},
  month = {Oct},
  publisher = {American Physical Society},
  doi = {10.1103/PhysRevApplied.8.044019},
  url = {https://link.aps.org/doi/10.1103/PhysRevApplied.8.044019}
}

@article{SWAIN2020101664,
title = {A feasibility study to measure magnetocardiography (MCG) in unshielded environment using first order gradiometer},
journal = {Biomedical Signal Processing and Control},
volume = {55},
pages = {101664},
year = {2020},
issn = {1746-8094},
doi = {https://doi.org/10.1016/j.bspc.2019.101664},
url = {https://www.sciencedirect.com/science/article/pii/S1746809419302459},
author = {Pragyna parimita Swain and S. Sengottuvel and Rajesh Patel and Awadhesh Mani and K. Gireesan}
}

@article{unshieldedopm,
  title = {A movable unshielded magnetocardiography system
},
  author = {Xiao, Wei and  Sun,Chenxi and  Shen,Liang and  Feng,Yulong and  Liu,Meng and  Wu,Yulong and  Liu,Xiyu and  Wu,Teng and  Peng,Xiang and  Guo,Hong },
  journal = {Science Advances },
  volume = {9},
  issue = {13},
  pages = {},
  numpages = {},
  year = {2023},
  month = {},
  publisher = {},
  doi = {10.1126/sciadv.adg1746},
  url = {https://www.science.org/doi/10.1126/sciadv.adg1746}
}

@article{Altarev2014MagneticallyShieldedRoom,
  title        = {A magnetically shielded room with ultra low residual field and gradient},
  author       = {Altarev, I. and Babcock, E. and Beck, D. and Burghoff, M. and Chesnevskaya, S. and Chupp, T. and Degenkolb, S. and Fan, I. and Fierlinger, P. and Frei, A. and Gutsmiedl, E. and Knappe-Gr{\"u}neberg, S. and Kuchler, F. and Lauer, T. and Link, P. and Lins, T. and Marino, M. and McAndrew, J. and Niessen, B. and Paul, S. and Petzoldt, G. and Schl{\"a}pfer, U. and Schnabel, A. and Sharma, S. and Singh, J. and Stoepler, R. and Stuiber, S. and Sturm, M. and Taubenheim, B. and Trahms, L. and Voigt, J. and Zechlau, T.},
  journal      = {Review of Scientific Instruments},
  volume       = {85},
  number       = {7},
  pages        = {075106},
  year         = {2014},
  doi          = {10.1063/1.4886146},
  url          = {https://doi.org/10.1063/1.4886146}
}

@article{lenz2021magnetic,
  author    = {Till Lenz and Arne Wickenbrock and Fedor Jelezko and Gopalakrishnan Balasubramanian and Dmitry Budker},
  title     = {Magnetic sensing at zero field with a single nitrogen‑vacancy center},
  journal   = {Quantum Science and Technology},
  year      = {2021},
  volume    = {6},
  number    = {4},
  pages     = {044010},
  doi       = {10.1088/2058-9565/abffbd},
  url       = {https://iopscience.iop.org/article/10.1088/2058-9565/abffbd},
  publisher = {IOP Publishing},
}

@article{PhysRevLett.126.197702,
  title = {Nuclear Spin Gyroscope based on the Nitrogen Vacancy Center in Diamond},
  author = {Soshenko, Vladimir V. and Bolshedvorskii, Stepan V. and Rubinas, Olga and Sorokin, Vadim N. and Smolyaninov, Andrey N. and Vorobyov, Vadim V. and Akimov, Alexey V.},
  journal = {Phys. Rev. Lett.},
  volume = {126},
  issue = {19},
  pages = {197702},
  numpages = {6},
  year = {2021},
  month = {},
  publisher = {American Physical Society},
  doi = {10.1103/PhysRevLett.126.197702},
  url = {https://link.aps.org/doi/10.1103/PhysRevLett.126.197702}
}

@article{substractionodmr,
  title = {Diamond quantum magnetometer with dc sensitivity of sub ten picotesla per squareroot Hz toward measurement of biomagnetic field

},
  author = { Sekiguchi,N. and Fushimi, M. and  Yoshimura,A. and   Shinei,C. and  Miyakawa, M. and  Taniguchi, T.
 Teraji, T. and  Abe, H. and Onoda, S. and
 Ohshima,T. and  Hatano,M. and Sekino, M. 
and  Iwasaki,T.},
  journal = {Phys. Rev. Applied},
  volume = {21},
pages = {064010 },
  numpages = {064010},
  year = {2024},
  publisher = {Phys. Rev. Applied },
  doi = {https://doi.org/10.1103/PhysRevApplied.21.064010},
  url = {https://journals.aps.org/prapplied/abstract/10.1103/PhysRevApplied.21.064010}
}

@article{Fenici2005ClinicalMCG,
  author  = {Fenici, Riccardo and Brisinda, Donatella and Meloni, Anna Maria},
  title   = {Clinical application of magnetocardiography},
  journal = {Expert Review of Molecular Diagnostics},
  year    = {2005},
  volume  = {5},
  number  = {3},
  pages   = {291--313},
  doi     = {10.1586/14737159.5.3.291},
  pmid    = {15934809},
}

@article{subpicowracht,
  title = {Diamond Magnetometry and Gradiometry Towards Subpicotesla dc Field Measurement},
  author = {Zhang, Chen and Shagieva, Farida and Widmann, Matthias and K\"ubler, Michael and Vorobyov, Vadim and Kapitanova, Polina and Nenasheva, Elizaveta and Corkill, Ruth and R\"ohrle, Oliver and Nakamura, Kazuo and Sumiya, Hitoshi and Onoda, Shinobu and Isoya, Junichi and Wrachtrup, J\"org},
  journal = {Phys. Rev. Applied},
  volume = {15},
  issue = {6},
  pages = {064075},
  numpages = {11},
  year = {2021},
  month = {Jun},
  publisher = {American Physical Society},
  doi = {10.1103/PhysRevApplied.15.064075},
  url = {https://link.aps.org/doi/10.1103/PhysRevApplied.15.064075}
}

@article{zerohuijie,
  title = {Zero-Field Magnetometry Based on Nitrogen-Vacancy Ensembles in Diamond},
  author = {Zheng, Huijie and Xu, Jingyan and Iwata, Geoffrey Z. and Lenz, Till and Michl, Julia and Yavkin, Boris and Nakamura, Kazuo and Sumiya, Hitoshi and Ohshima, Takeshi and Isoya, Junichi and Wrachtrup, J\"org and Wickenbrock, Arne and Budker, Dmitry},
  journal = {Phys. Rev. Appl.},
  volume = {11},
  issue = {6},
  pages = {064068},
  numpages = {7},
  year = {2019},
  month = {Jun},
  publisher = {American Physical Society},
  doi = {10.1103/PhysRevApplied.11.064068},
  url = {https://link.aps.org/doi/10.1103/PhysRevApplied.11.064068}
}

@article{2019BarryNVReview,
  title = {Sensitivity optimization for NV-diamond magnetometry},
  author = {Barry, J. F. and Schloss, J. M. and Bauch, E. and Turner, M. J. and Hart, C. A. and Pham, L. M. and Walsworth, R. L.},
  journal = {Rev. Mod. Phys.},
  volume = {92},
  issue = {1},
  pages = {015004},
  numpages = {68},
  year = {2020},
  month = {Mar},
  publisher = {American Physical Society},
  doi = {10.1103/RevModPhys.92.015004},
  url = {https://link.aps.org/doi/10.1103/RevModPhys.92.015004}
}

@article{HeartRateVariability,
    author = {},
    title = {Heart rate variability. {S}tandards of measurement, physiolog-ical interpretation, 
            and clinical use. {T}ask for of the {E}uropean{S}ociety of {C}ardiology and the {N}orth {A}merican {S}ociety of {P}acing and {E}lectrophysiology},
    journal = {Eur Heart J},
    volume = {17},
    pages = {354-81},
    year = {1996}
}

@article{Heliyon2024Su,
author = {Shengran Su and Zhenyuan Xu and Xiang He and Guoyi Zhang and Haijun Wu and Yalan Gao and Yueliang Ma and Chanling Yin and Yi Ruan and Kan Li and Qiang Lin},
title = {Vector magnetocardiography using compact optically-pumped magnetometers},
journal = {Heliyon},
volume = {10},
number = {7},
pages = {e29092},
year = {2024},
doi = {https://doi.org/10.1016/j.heliyon.2024.e29092},
}

@article{Zheng2020OPG,
title = {Vector magnetocardiography measurement with a compact elliptically polarized laser-pumped magnetometer},
author = {Wenqiang Zheng and Shengran Su and Guoyi Zhang and Xin Bi and Qiang Lin},
journal = {Biomed. Opt. Express},
number = {2},
pages = {649--659},
publisher = {Optica Publishing Group},
volume = {11},
month = {Feb},
year = {2020},
url = {https://opg.optica.org/boe/abstract.cfm?URI=boe-11-2-649},
doi = {10.1364/BOE.380314},
}

@article{YoungJin2019APL,
    author = {Kim, Young Jin and Savukov, Igor and Newman, Shaun},
    title = {Magnetocardiography with a 16-channel fiber-coupled single-cell Rb optically pumped magnetometer},
    journal = {Applied Physics Letters},
    volume = {114},
    number = {14},
    pages = {143702},
    year = {2019},
    month = {04},
    doi = {10.1063/1.5094339},
    url = {https://doi.org/10.1063/1.5094339},
}

@article{eisenach2021cavity,
  title={Cavity-enhanced microwave readout of a solid-state spin sensor},
  author={Eisenach, Erik R and Barry, John F and O’Keeffe, Michael F and Schloss, Jennifer M and Steinecker, Matthew H and Englund, Dirk R and Braje, Danielle A},
  journal={Nature communications},
  volume={12},
  number={1},
  pages={1357},
  year={2021},
  publisher={Nature Publishing Group UK London},
 doi = {https://doi.org/10.1038/s41467-021-21256-7},
}

@article{wang2025exceptional,
  title={Exceptional sensitivity near the bistable transition point of a hybrid quantum system},
  author={Wang, Hanfeng and Jacobs, Kurt and Fahey, Donald and Hu, Yong and Englund, Dirk R and Trusheim, Matthew E},
  journal={arXiv preprint arXiv:2507.09691},
  year={2025}
}

@article{wu2025spin,
  title={Spin squeezing in an ensemble of nitrogen-vacancy centers in diamond},
  author={Wu, Weijie and Davis, Emily J and Hughes, Lillian B and Ye, Bingtian and Wang, Zilin and Kufel, Dominik and Ono, Tasuku and Meynell, Simon A and Block, Maxwell and Liu, Che and others},
  journal={arXiv preprint arXiv:2503.14585},
  year={2025}
}

@article{gao2025signal,
  title={Signal amplification in a solid-state quantum sensor via asymmetric time-reversal of many-body dynamics},
  author={Gao, Haoyang and Martin, Leigh S and Hughes, Lillian B and Leitao, Nathaniel T and Put, Piotr and Zhou, Hengyun and Koyluoglu, Nazli U and Meynell, Simon A and Jayich, Ania C Bleszynski and Park, Hongkun and others},
  journal={arXiv preprint arXiv:2503.14598},
  year={2025}
}

@article{fescenko2020diamond,
  title={Diamond magnetometer enhanced by ferrite flux concentrators},
  author={Fescenko, Ilja and Jarmola, Andrey and Savukov, Igor and Kehayias, Pauli and Smits, Janis and Damron, Joshua and Ristoff, Nathaniel and Mosavian, Nazanin and Acosta, Victor M},
  journal={Physical review research},
  volume={2},
  number={2},
  pages={023394},
  year={2020},
  publisher={APS},
  doi = {https://doi.org/10.1103/PhysRevResearch.2.023394},
}

@article{xie2021hybrid,
  title={A hybrid magnetometer towards femtotesla sensitivity under ambient conditions},
  author={Xie, Yijin and Yu, Huiyao and Zhu, Yunbin and Qin, Xi and Rong, Xing and Duan, Chang-Kui and Du, Jiangfeng},
  journal={Science Bulletin},
  volume={66},
  number={2},
  pages={127--132},
  year={2021},
  publisher={Elsevier},
  url = {https://www.sciengine.com/SB/doi/10.1016/j.scib.2020.08.001},
  doi = {https://doi.org/10.1016/j.scib.2020.08.001},
}

@article{newman2025endoscopic,
  title={Endoscopic fiber-coupled diamond magnetometer for cancer surgery},
  author={Newman, AJ and Graham, SM and Stephen, CJ and Edmonds, AM and Markham, ML and Morley, GW},
  journal={arXiv preprint arXiv:2504.05884},
  year={2025}
}

@inbook{Taulu2014NovelNoiseReduction,
  author       = {Samu Taulu and Juha Simola and Jukka Nenonen and Lauri Parkkonen},
  title        = {Novel Noise Reduction Methods},
  publisher    = {Springer},
  pages        = {35-71},
  doi          = {10.1007/978-3-642-33045-2_2},
  url          = {https://link.springer.com/chapter/10.1007/978-3-642-33045-2_2}
}

@article{Kiehl2024heading,
  author       = {C. Kiehl and T. S. Menon and D. P. Hewatt and S. Knappe and T. Thiele and C. A. Regal},
  title        = {Correcting heading errors in optically pumped magnetometers through microwave interrogation},
  journal      = {Physical Review Applied},
  volume       = {22},
  number       = {1},
  pages        = {014005},
  year         = {2024},
  doi          = {10.1103/PhysRevApplied.22.014005},
  issn         = {2331-7019},
  publisher    = {American Physical Society},
}

@article{Tian2024DeadZone,
  author       = {Mengnan Tian and Liwei Jiang and Xin Zhao and Yanchao Chai and Jiali Liu and Zhenglong Lu and Wei Quan},
  title        = {Dead-zone suppression method of NMOR atomic magnetometers based on alignment and orientation polarization},
  journal      = {Sensors and Actuators A: Physical},
  volume       = {379},
  pages        = {115842},
  year         = {2024},
  doi          = {10.1016/j.sna.2024.115842},
  url          = {https://www.sciencedirect.com/science/article/pii/S0924424724008367}
}

@article{Griffith2009Miniature,
  author  = {W. C. Griffith and Ricardo Jimenez-Martinez and Vishal Shah and Svenja Knappe and John Kitching},
  title   = {Miniature atomic magnetometer integrated with flux concentrators},
  journal = {Applied Physics Letters},
  volume  = {94},
  number  = {2},
  pages   = {023502},
  year    = {2009},
  doi     = {10.1063/1.3056152},
  url     = {https://doi.org/10.1063/1.3056152}
}

@article{Strasburger2008Magnetocardiography,
  author  = {Strasburger, Janette F. and Cheulkar, Bageshree and Wakai, Ronald T.},
  title   = {Magnetocardiography for fetal arrhythmias},
  journal = {Heart Rhythm},
  year    = {2008},
  volume  = {5},
  number  = {7},
  pages   = {1073--1076},
  doi     = {10.1016/j.hrthm.2008.02.035},
  pmid    = {18486565},
  pmcid   = {PMC2574560},
}

@misc{DestatisCostOfIllness,
  author       = {{German Federal Statistical Office (Destatis)}},
  title        = {Cost of Illness},
  url          = {https://www.destatis.de/EN/Press/2017/09/PE17_347_236.html},
  publisher    = {Federal Statistical Office (Destatis), Germany},
  year         = {2017},
  note         = {Accessed: 2025-12-23},
}

\end{document}